\documentclass[twoside,twocolumn,9pt]{article}
\usepackage{extsizes}
\usepackage[super,sort&compress,comma]{natbib} 
\usepackage[version=3]{mhchem}
\usepackage[left=1.5cm, right=1.5cm, top=1.785cm, bottom=2.0cm]{geometry}
\usepackage{balance}
\usepackage{times,mathptmx}
\usepackage{sectsty}
\usepackage{graphicx} 
\usepackage{lastpage}
\usepackage[format=plain,justification=justified,singlelinecheck=false,font={stretch=1.125,small,sf},labelfont=bf,labelsep=space]{caption}
\usepackage{float}
\usepackage{fancyhdr}
\usepackage{fnpos}
\usepackage[english]{babel}
\usepackage{array}
\usepackage{droidsans}
\usepackage{charter}
\usepackage[T1]{fontenc}
\usepackage[usenames,dvipsnames]{xcolor}
\usepackage{setspace}
\usepackage[compact]{titlesec}

\usepackage{amssymb}
\usepackage{float}
\usepackage{subcaption}
\usepackage{cases}
\usepackage{gensymb}

\usepackage{epstopdf}

\definecolor{cream}{RGB}{222,217,201}

\begin{document}

\pagestyle{fancy}
\thispagestyle{plain}
\fancypagestyle{plain}{

\fancyhead[C]{\includegraphics[width=18.5cm]{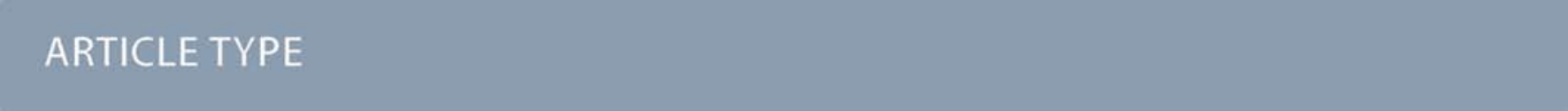}}
\fancyhead[L]{\hspace{0cm}\vspace{1.5cm}\includegraphics[height=30pt]{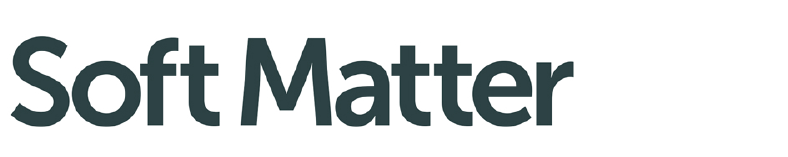}}
\fancyhead[R]{\hspace{0cm}\vspace{1.7cm}\includegraphics[height=55pt]{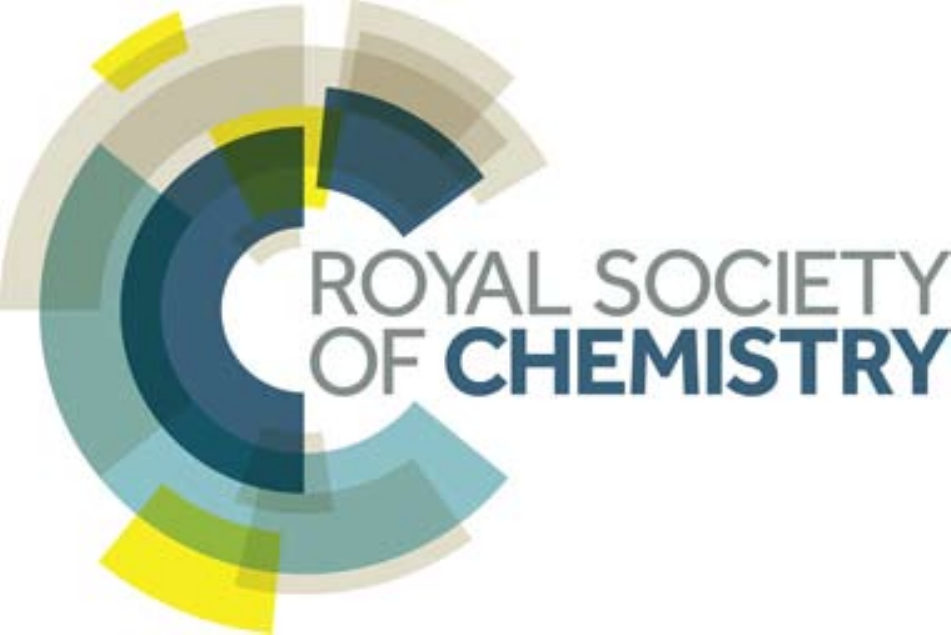}}
\renewcommand{\headrulewidth}{0pt}
}

\makeFNbottom
\makeatletter
\renewcommand\LARGE{\@setfontsize\LARGE{15pt}{17}}
\renewcommand\Large{\@setfontsize\Large{12pt}{14}}
\renewcommand\large{\@setfontsize\large{10pt}{12}}
\renewcommand\footnotesize{\@setfontsize\footnotesize{7pt}{10}}
\makeatother

\renewcommand{\thefootnote}{\fnsymbol{footnote}}
\renewcommand\footnoterule{\vspace*{1pt}%
\color{cream}\hrule width 3.5in height 0.4pt \color{black}\vspace*{5pt}} 
\setcounter{secnumdepth}{5}

\makeatletter 
\renewcommand\@biblabel[1]{#1}            
\renewcommand\@makefntext[1]%
{\noindent\makebox[0pt][r]{\@thefnmark\,}#1}
\makeatother 
\renewcommand{\figurename}{\small{Fig.}~}
\sectionfont{\sffamily\Large}
\subsectionfont{\normalsize}
\subsubsectionfont{\bf}
\setstretch{1.125} 
\setlength{\skip\footins}{0.8cm}
\setlength{\footnotesep}{0.25cm}
\setlength{\jot}{10pt}
\titlespacing*{\section}{0pt}{4pt}{4pt}
\titlespacing*{\subsection}{0pt}{15pt}{1pt}

\fancyfoot{}
\fancyfoot[LO,RE]{\vspace{-7.1pt}\includegraphics[height=9pt]{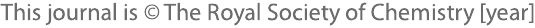}}
\fancyfoot[CO]{\vspace{-7.1pt}\hspace{13.2cm}\includegraphics{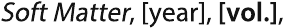}}
\fancyfoot[CE]{\vspace{-7.2pt}\hspace{-14.2cm}\includegraphics{RF}}
\fancyfoot[RO]{\footnotesize{\sffamily{1--\pageref{LastPage} ~\textbar  \hspace{2pt}\thepage}}}
\fancyfoot[LE]{\footnotesize{\sffamily{\thepage~\textbar\hspace{3.45cm} 1--\pageref{LastPage}}}}
\fancyhead{}
\renewcommand{\headrulewidth}{0pt} 
\renewcommand{\footrulewidth}{0pt}
\setlength{\arrayrulewidth}{1pt}
\setlength{\columnsep}{6.5mm}
\setlength\bibsep{1pt}

\makeatletter 
\newlength{\figrulesep} 
\setlength{\figrulesep}{0.5\textfloatsep} 

\newcommand{\topfigrule}{\vspace*{-1pt}%
\noindent{\color{cream}\rule[-\figrulesep]{\columnwidth}{1.5pt}} }

\newcommand{\botfigrule}{\vspace*{-2pt}%
\noindent{\color{cream}\rule[\figrulesep]{\columnwidth}{1.5pt}} }

\newcommand{\dblfigrule}{\vspace*{-1pt}%
\noindent{\color{cream}\rule[-\figrulesep]{\textwidth}{1.5pt}} }

\makeatother

\newcommand{\eqv}[1]{\begin{equation} #1 \veq \end{equation}}
\newcommand{\eqp}[1]{\begin{equation} #1 \peq \end{equation}}
\newcommand{\eq}[1]{\begin{equation} #1  \end{equation}}
\newcommand{\und}[1]{\underline{ #1 }}
\newcommand{\undd}[1]{\underline{\underline{ #1 }}}
\newcommand{\dd}[2]{\dfrac{\mathrm d \,#1}{\mathrm d\,  #2}}
\newcommand{\ddp}[2]{\dfrac{\partial \, #1}{\partial \, #2}}
\newcommand{\pic}[3]{\begin{figure}[h!]\begin{center} \includegraphics[width=#2]{#1}\caption{#3 } \end{center}\end{figure}}
\newcommand{\ppar}[1]{\left( #1 \right)}
\newcommand{\un}[1]{\,\mathrm{#1}}
\newcommand{\del}{\mathrm{d}}
\newcommand{\inv}[1]{\dfrac{1}{#1}}
\newcommand{\ind}[1]{_{\mathrm{#1}}}
\newcommand{\ej}{\mathrm{e}^{\, j\omega t}}
\newcommand{\abr}[3]{$#1$ & $\mathrm{#2}$ & #3 \\}
\newcommand{\kt}{k\ind{B}T}
\newcommand{\paa}[2]{\mbox{PAA-$#1$-C$\ind{#2}$}}
\newcommand{\fig}[1]{Figure~\ref{#1}}
\newcommand{\angstrom}{\mbox{\normalfont\AA}}
\newcommand{\mum}{\, \mathrm{\mu m}}
\newcommand{\mnm}{\, \mathrm{mN/m}}
\newcommand{\etal}{\textit{et al.}}
\newcommand{\pd}[1]{\cdot 10^{#1}}
\newcommand{\idT}{\undd{\mathds{1}}}
\newcommand{\peq}{\; \; .}
\newcommand{\veq}{\; \; ,}

\newcommand{\eqcv}[3]{\begin{subnumcases}{#1}
\vphantom{\Bigg(}#2 \veq \\ 
\vphantom{\Bigg(}#3 \veq
\end{subnumcases}} 

\newcommand{\eqcp}[3]{\begin{subnumcases}{#1}
\vphantom{\Bigg(}#2 \veq \\
\vphantom{\Bigg(}#3 \peq
\end{subnumcases}}


\twocolumn[
  \begin{@twocolumnfalse}
\vspace{3cm}
\sffamily
\begin{tabular}{m{4.5cm} p{13.5cm} }

\includegraphics{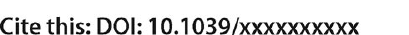} & \noindent\LARGE{\textbf{Microfluidic probing of the complex interfacial rheology of multilayer capsules.}} \\
\vspace{0.3cm} & \vspace{0.3cm} \\

 & \noindent\large{Corentin Tr{\'e}gou{\"{e}}t\textit{$^{a}$}\textit{$^{b}$}, Thomas Salez \textit{$^{c}$}\textit{$^{d}$}, C\'ecile Monteux \textit{$^{b}$}\textit{$^{d}$} and Mathilde Reyssat \textit{$^{a}$}} \\

\includegraphics{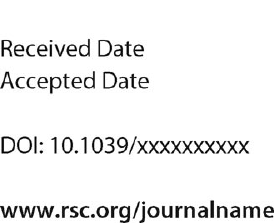} & \noindent\normalsize{Encapsulation of chemicals using polymer membranes enables to control their transport and delivery for applications such as agrochemistry or detergency. To rationalize the design of polymer capsules, it is necessary to understand how the membranes' mechanical properties control the transport and release of the cargo. In this article, we use microfluidics to produce model polymer capsules and study in situ their behavior in controlled elongational flows. Our model capsules are obtained by assembling polymer mono and hydrogen-bonded bilayers at the surface of an oil droplet in water. We also use microfluidics to probe in situ the mechanical properties of the membranes in a controlled elongational flow generated by introducing the capsules through a constriction and then in a larger chamber. The deformation and relaxation of the capsules depend on their composition and especially on the molecular interactions between the polymer chains that form the membranes and the anchoring energy of the first layer. We develop a model and perform numerical simulations to extract the main interfacial properties of the capsules from the measurement of their deformations in the microchannels.} \\

\end{tabular}

\end{@twocolumnfalse} \vspace{0.6cm}

]

\renewcommand*\rmdefault{bch}\normalfont\upshape
\rmfamily
\section*{}
\vspace{-1cm}


\footnotetext{\textit{$^{a}$~UMR CNRS Gulliver 7083, ESPCI Paris, PSL Research University, 75005 Paris, France. E-mail: mathilde.reyssat@espci.fr}}
\footnotetext{\textit{$^{b}$~UMR CNRS SIMM 7615, ESPCI Paris, PSL Research University, 75005 Paris, France. E-mail: cecile.monteux@espci.fr}}
\footnotetext{\textit{$^{c}$~Univ. Bordeaux, CNRS, LOMA, UMR 5798, F-33405 Talence, France. }}
\footnotetext{\textit{$^{d}$~Global Station for Soft Matter, Global Institution for Collaborative Research and Education, Hokkaido University, Sapporo, Japan.}}

\footnotetext{\dag~Electronic Supplementary Information (ESI) available. See DOI: 10.1039/cXsm00000x/}



\section{Introduction}

Droplets with valuable content can be protected from their environment by a membrane, usually elastic, which turns them into capsules. Encapsulation of droplets allows a better control on the delivery of the droplet content. Capsules are used in a very broad range of applications such as food industry \cite{Klinkesorn2005}, drug delivery \cite{Delcea2011,Correa2016}, material science \cite{Jinglei2008} or cosmetics \cite{LeTirilly2015}. Various types of processes have been developed in the last decades to produce capsules, especially thanks to the development of microfluidic technics \cite{Gao2001a,Antipov2001, Priest2008,Kantak2011, Erni2011, Polenz2015b, Bjornmalm2015, DupredeBaubigny2017}.

To rationalize the formulation of capsules it is necessary to understand the link between the composition of the protecting membrane, its mechanical properties and the  performances of the capsules, which requires model polymer membranes as well as experimental techniques to determine capsule mechanical properties.
New experimental techniques have been developed to characterize elastic polymer membranes at liquid interfaces \cite{Vandebril2009,Sarrazin2016,Xie2017,Knoche2013,Oshri2015}. The rheology of such membranes has been studied in model geometries \cite{Gordon2004, Carvajal2011, Erni2012, Ferri2012, Knoche2013, Kleinberger2013,Sarrazin2016} and also on real capsules in viscous flows to mimic real conditions of fabrication and use, through various experiments \cite{Walter2001, Prevot2003, Hu2013, Kleinberger2013,Xie2017}, theories  \cite{BarthesBiesel1981, Finken2011, Higley2012} and simulations \cite{Pozrikidis1995, Park2013, Koolivand2017}. From this perspective, microfluidics is an ideal tool which allows to integrate measurement units directly \textit{in situ} in the devices and to perform automated measurements on a large number of capsules. 

The use of extensional flows to measure the surface properties of droplets has been first developed by Taylor \cite{Taylor1934} with the so-called \textit{four-roller apparatus}. Since then, this method has been adapted in microfluidics by imposing an extensional flow field to droplets at a transition between a narrow channel and a larger chamber, and has been firstly employed to measure interfacial tensions \cite{Cabral2006,Tregouet2018-PRF}. In our previous work \cite{Tregouet2018-PRF}, by simulating the flow field in the large chamber we were able to determine the shear forces applying on the capsules. Assuming a competition between the viscous forces deforming the capsules and the interfacial tension force restoring its shape, we were then able to predict the shape of the droplet as a function of its position in the channel for a given interfacial tension. Inversely, the fit of our experimental data with the model enabled to deduce the interfacial tension of the droplets. \\Recently, similar  design has been used to measure the elastic modulus of capsules \cite{Leclerc2011, Polenz2015, Gires2016}. However, for intermediate situations where several interfacial components can interfere (for example in the case of visco-elastic capsules for which the interfacial tension force is not negligible), there is no model which can rationalize the evolution of the capsules deformation over time by accounting for the relative importances of membrane viscoelastic properties and interfacial tension.\\

In previous studies published in our group \cite{LeTirilly2015,LeTirilly2016a,Tregouet2016}, we have designed model polymer membranes with controlled interfacial properties. These membranes are obtained by assembling polymer multilayers at oil water interfaces through hydrogen bonds. We have shown that the hydrophobicity of the polymers enables to control the interaction energy of the assembled layers by an interplay of hydrogen bonds and hydrophobic interactions, which tunes the shear moduli of the membrane. Moreover, controlling the anchoring energy of the chains by adding hydrophobic anchors on the chains enables to control the compression moduli of the adsorbed layers. 

In the present article, our goal is to study the relation between the composition of the polymer membranes protecting the droplets and the deformation of capsules in a controlled flow field. We produce, using microfluidics, monodisperse populations of oil droplets covered by polymer multilayers of various compositions described above and we study their deformation in microfluidic extensional flows, obtained by the combination of a constriction followed by a larger chamber. We find that the dynamics of the capsules under this extensional flow strongly depends on the composition of the multilayers. We propose a coarse-grained model to extract the different characteristic properties of each polymer membrane, and evaluate the relative importances of the various factors ruling the capsule's deformation.

\section{Materials and methods}
Oil droplets in aqueous solutions are covered with polymers in a microfluidic chip, and directed through an extensional flow to measure their deformation and relaxation.

\subsection{Materials}

\subsubsection{Liquids and polymers}

The droplets consist of mineral oil (Sigma Aldrich,  Mineral Oil Rotational Viscosity Standard of viscosity $29.04 \un{mPa.s}$ at $25.00 \un{\degree C}$) flowing in different polymer solutions, prepared at $1 \un{\%w}$ and adjusted at pH $3$ with molar solutions of HCl or NaOH.\\
In recent articles \cite{LeTirilly2015,LeTirilly2016a} we have studied the interfacial rheological properties of polymer multi layers assembled at oil water interfaces by hydrogen bonds. 
The polymers used to obtain the multilayers are:
\begin{itemize}
\item poly(methacrylic acid), (PMAA) provided by Polysciences, $M\ind{w}=50 000 \un{g/mol}$,
\item poly(vinyl pyrrolidone), (PVP) provided by Fluka, $M\ind{w}=40 000 \un{g/mol}$, 
\item covalently randomly hydrophobically modified poly(acrylic acid) \paa{0.7}{12} and \paa{0.8}{8}: respectively $0.7 \%$ and $0.8 \%$ of the monomers are grafted to an alkyl chain of 12 and 8 carbons respectively, the backbone being provided by Polysciences Inc., $M\ind{w}=100 000 \un{g/mol}$. We prepared these polymers by following the protocol of Wang \etal \cite{Wang1988}. The grafting density has been measured by proton NMR.
\end{itemize}

The droplets are characterized in the polymer solution leading to the last adsorbed layer. 

\subsubsection{Microfluidic chips}
The droplets are produced by microfluidics using a standard flow-focusing device made of poly(dimethyl siloxane) (PDMS) in the production chip \cite{Tregouet2018-PRF}. Such a setup provides a very monodisperse collection of droplets, tunable from 50 to 70 $\mum$ in diameter depending on the experimental conditions. As described in the following, we design specific connections to add a second polymer layer. The production chip is further connected to a characterization chip realized with a photosensitive adhesive provided by Norland (NOA 81). This microfabrication technique is a way to build rigid channels confined between two glass surfaces, and able to sustain large flow rates without being deformed~\cite{Bartolo2008}. The inlet of the chip is connected to the droplet-production chip by a silicon tubing (Tygon) of inner diameter $800 \mum$, coated with a solution of Bovine Serum Albumine (BSA) (Sigma Aldrich) to prevent adhesion of the droplets inside the tubing (incubation at $1 \un{\%w}$ solution during $12 \un{h}$ at room temperature), whereas its outlet is connected to a Peek tubing of inner diameter $125 \mum$.

\subsection{Methods}

\subsubsection{Fabrication of multilayers}

\begin{figure}[ht!]
	\centering
\includegraphics[width= 0.9 \linewidth]{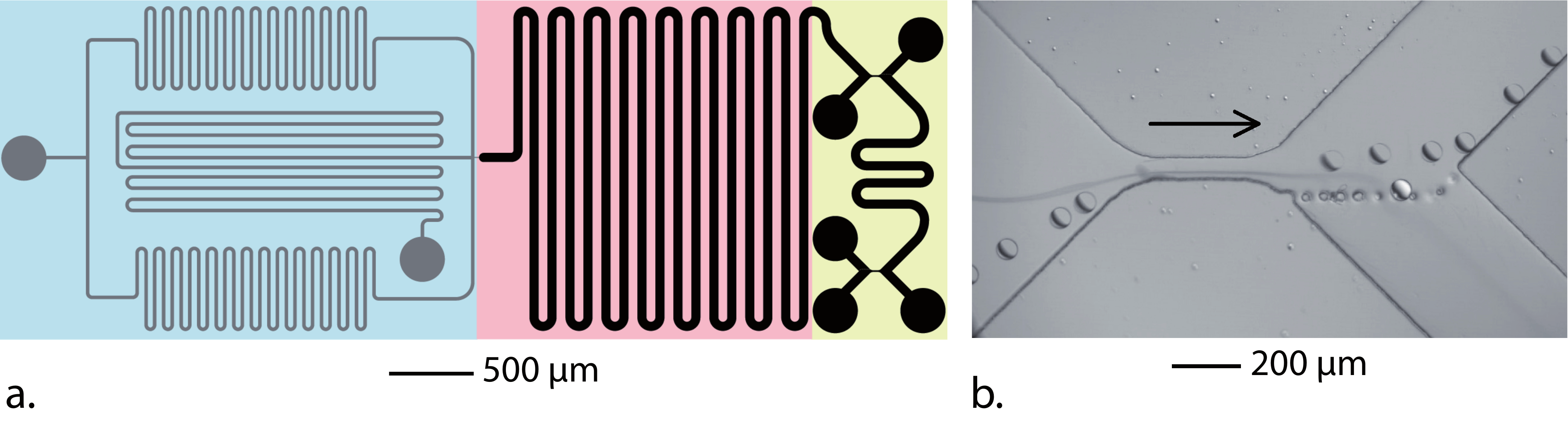}
\caption{a. Design of the capsule-production microfluidic chip. The blue part corresponds to the flow focusing. The serpentines at the entrance of the fluids increase the resistance of the channels and give a better control of the stability of the droplets. The red part corresponds to an incubation channel, whereas the yellow one is made of two successive phase changes. b: zoom on a phase change set-up made of a pinched flow fractionation device used twice, to transfer the droplet from the first polymer solution to the rinsing solution, and then from the rising solution to a second polymer solution. In each case, the two solutions are miscible but have different optical indices, which makes the frontier visible using digital interferometric contrast. On this picture, the monodisperse droplets coming from the bottom left cross the confined interface and reach the top right channel. A barrier formed by a row of small posts prevents the droplets from going to the bottom right channel.}
        \label{PFF}
\end{figure}

The set-up is sketched in \fig{PFF}. Our first objective is to build capsules with a multilayer membrane. Once the droplets are produced \cite{Tregouet2018-PRF}, they are forced to circulate into a long serpentine to let the first polymer adsorb at their surfaces \cite{Priest2008}. The addition of a second polymer layer is done in two steps. First, the droplets are transferred to a solution of HCl (pH 3) through a pinched-flow fractionation as described by Yamada \etal  \cite{Yamada2004} and illustrated in \fig{PFF} b. Secondly, using the same method, the droplets are transferred from the HCl solution to a second polymer solution.

\subsubsection{Capsule deformation}

\begin{figure}[ht!]
	\begin{center}
		\includegraphics[width=0.45\textwidth]{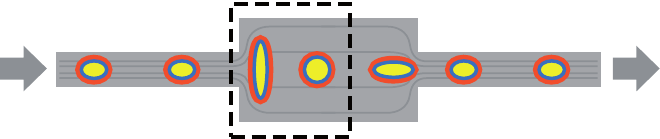}
		\caption{\label{SchematicDeformation} Schematic of the characterization chip. The deformation, initially negative when the capsules are confined in the channel, becomes positive when they enter the chamber, and finally relaxes towards zero (spherical shape) in the middle of the chamber. The dashed rectangle represents the observation frame.}
	\end{center}
\end{figure}

The principles of capsule deformation and observation were described previously \cite{Tregouet2018-PRF}. Briefly, the characterization chip (\fig{SchematicDeformation}) consists in a sharp transition between a narrow channel (of width $W=40 \mum$) and a wide chamber (of width $3 W$), similar to the chip used by Polenz \textit{et al.}~\cite{Polenz2015}. The height of the chamber (in the $z$-direction) is $100 \mum$: the droplets of diameter comprised between 50 and 70 $\mum$ are therefore not confined vertically in the chamber.

The divergent flow at the entrance of the wide channel imposes a high viscous stress on the capsules that elongate perpendicularly to the main direction of the flow. The entrance is therefore the region where we measure the deformation and the relaxation of the capsules. The flow is controlled by a pressure controller Fluigent.

The droplet deformation is defined as a dimensionless number calculated from the $x$-axis width $a$ and the $y$-axis height $b$ of the droplet, through:
\eqp{D=\dfrac{b-a}{b+a}\label{Dab}}
The deformation and speed are measured for a large number of identical capsules at different positions, which correspond to different times. Such a process allows for a precise measurement thanks to statistical analysis. Indeed, for each set, we measure about $8$ points per capsule, for about $25$ capsules.

The experimental evolution of the deformation as a function of time is then interpreted through simulations and a theoretical model which enable the extraction of the interfacial rheological properties. These properties are finally compared with values obtained from independent techniques: the pendant drop method and the interfacial shear rheometer.

\subsubsection{Simulations}

The details of the simulations used to calculate the shear stress applied by the surrounding fluid on the capsule are provided in a previous article \cite{Tregouet2018-PRF}. Briefly, using COMSOL we simulate the laminar flow in the chamber for several, fixed, positions of the capsule. The boundary condition at the interface is chosen such that the fluid velocity matches the experimentally-measured capsule velocity. The viscous stress around the capsule is then calculated and its main components are averaged on all the points in the close vicinity of the interface.

\subsubsection{Pendant drop}

To determine independently the surface tension of the droplets, we use the pendant-drop method (Tracker apparatus, Teclis, France). Details concerning the principle of this technique can be found in an article from Rotenberg et al. \cite{Rotenberg1983}. Briefly, a fresh millimetric oil drop is formed in the polymer solution and the surface tension is then obtained through the analysis of its shape.

\section{Results and discussion}

\subsection{Experimental results}

We observe that all the capsules are initially negatively deformed ($D<0$) as they exit the constriction due to the lateral confinement of the narrow channel. As they progress in the large chamber, their behavior depends on the nature of the interface. We stress that, in the following, the positions along the transport axis ($x$) and time ($t$) are equivalent, since the capsules' trajectories are recorded.

\begin{figure}[ht!]
	\begin{center}
		\hfill
		\begin{subfigure}[t]{0.45 \textwidth}
			\begin{center}
				\includegraphics[width=8 cm]{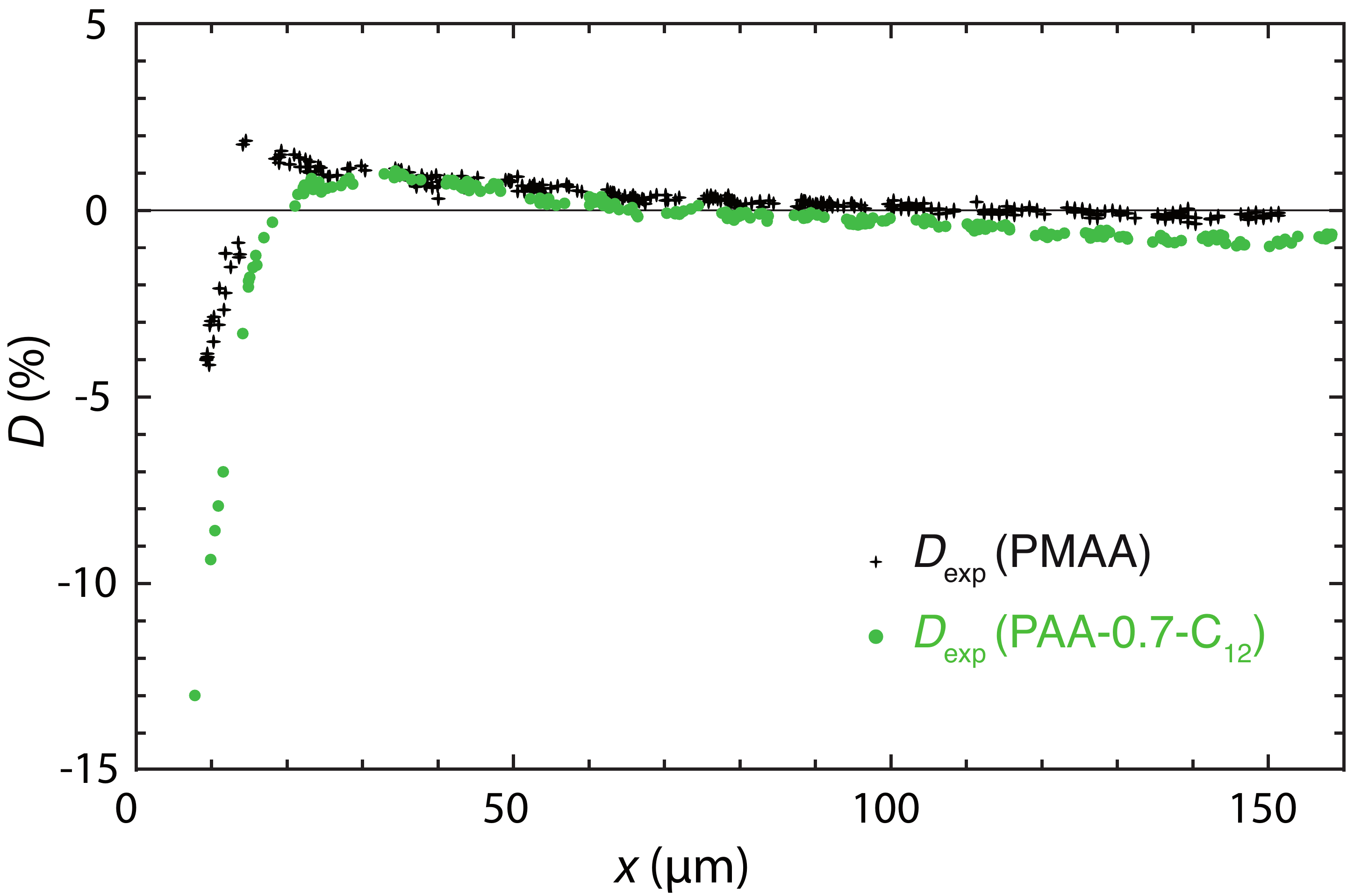}
				\caption{\label{DeformationMonolayers} Deformation as a function of position of the center of mass, for capsules made of PMAA and \paa{0.7}{12}, as indicated. The origin of $x$ corresponds to the entrance of the wide chamber. The PMAA data was previously studied~\cite{Tregouet2018-PRF} and serves here as a reference.}
			\end{center}
		\end{subfigure}
		\hfill
		\begin{subfigure}[t]{0.45 \textwidth}
			\begin{center}
				\includegraphics[width=8cm]{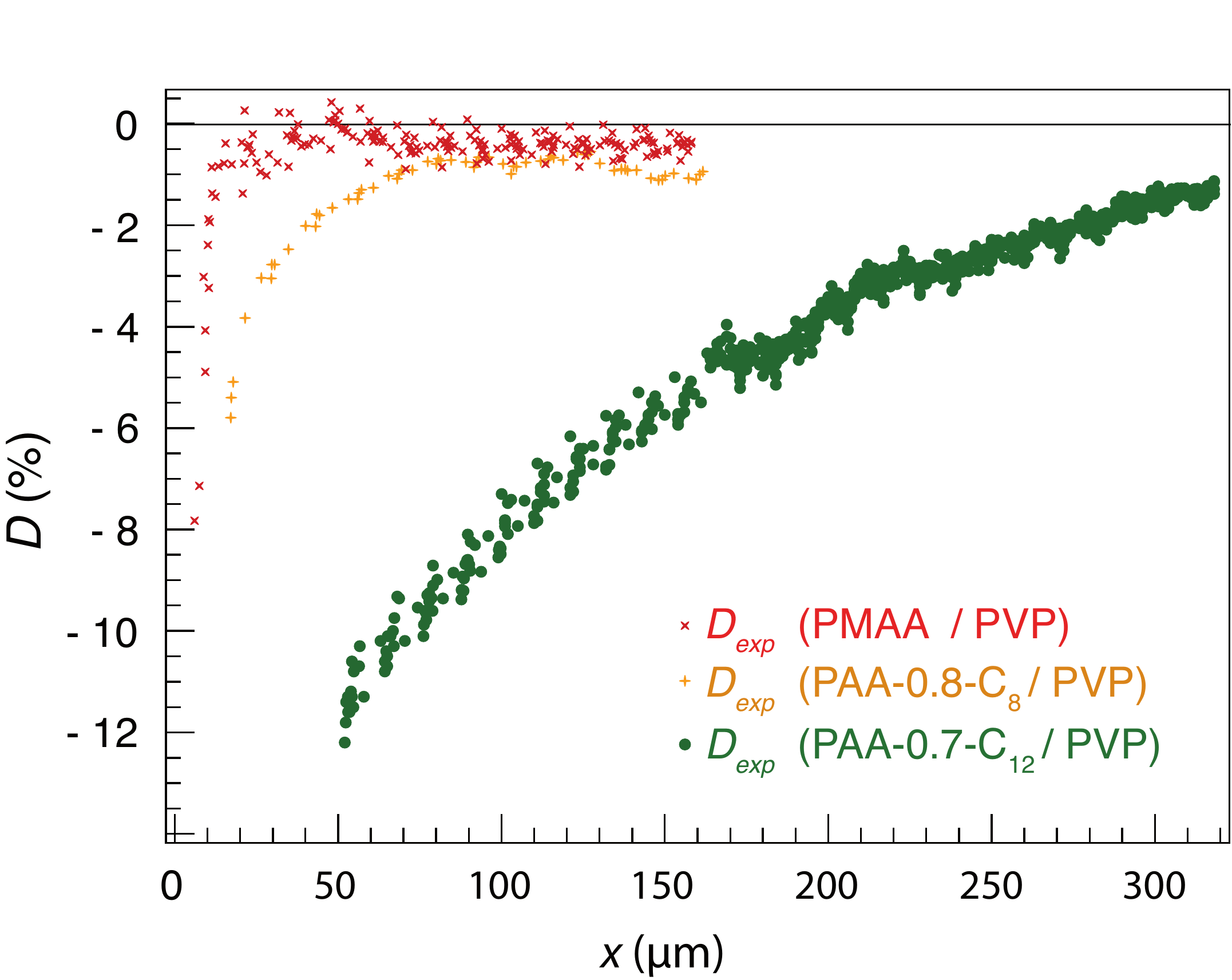}
				\caption{\label{DeformationPmaaLayers} Deformation of multilayer capsules versus position in the chamber, for various membrane compositions as indicated.}
			\end{center}
		\end{subfigure}
		\caption{\label{AnchoringEnergyMicroflu} Deformation of capsules made of different kinds of membrane.}
		\hfill \\
	\end{center}
\end{figure}

The case of the PMAA monolayers (\fig{DeformationMonolayers}) was already presented and analyzed theoretically in a previous article \cite{Tregouet2018-PRF}. In this case, the deformation quickly becomes positive at the constriction exit because of the elongational flow and then relaxes toward zero when the elongational stress decreases.

In the case of the monolayers of \paa{0.7}{12}, the deformation initially negative becomes positive, reaches a maximum, before becoming negative again (\fig{DeformationMonolayers}), unlike the PMAA monolayers.

We observe more pronounced qualitative differences for the two other types of membrane, as shown in \fig{DeformationPmaaLayers}: the deformation remains negative all along the trajectory and shows no maximum for the bilayers PMAA/PVP, \paa{0.7}{12}/PVP and \paa{0.8}{8}/PVP. In addition, we observe that \paa{0.7}{12}/PVP capsules relax significantly more slowly than all the other systems.

\subsection{Theoretical Analysis}

\subsubsection{Capillary model}

In a given flow field, a droplet adopts a steady deformation, $D\ind{steady}$ which was shown by Barth\`{e}s-Biesel \textit{et al.}~ \cite{BarthesBiesel1973} to depend on the viscosity ratio between the droplet and the surrounding phase $\lambda=\frac{\eta\ind{droplet}}{\eta\ind{bulk}}$, the radius of the droplet $r$, and the interfacial tension $\gamma$ between the two phases as follows\cite{BarthesBiesel1973} : 

\eqv{D\ind{steady}=\dfrac{19 \lambda +16}{16 \lambda +16} \cdot \dfrac{\eta\ind{bulk} \, r}{\gamma}\cdot \left(e\ind{max}-e\ind{min} \right)
\label{Dsteady}} where $e\ind{max}$ and $e\ind{min}$ are the eigenvalues of the deformation-rate tensor.

The same authors also described the transient regime leading to this steady deformation as a first-order relaxation in time~\cite{BarthesBiesel1973}. The derivative $\dot{D}$ of the deformation $D$ with respect to time $t$ thus depends on the steady deformation $D\ind{steady}$ and the actual deformation, through:
\eqv{\dot{D}=\inv{\tau\ind{ca}}\cdot \left(D\ind{steady}-D\right)\label{TransientD}}
with a relaxation time $\tau\ind{ca}$ defined as follows~\cite{BarthesBiesel1973}:
\eqp{\tau\ind{ca}=\dfrac{2}{5}\cdot (2 \lambda+3) \cdot\dfrac{19 \lambda +16}{16 \lambda +16} \cdot \dfrac{\eta\ind{bulk} \, r}{\gamma}\label{LiquidRelaxationTimeCapsule}}

\begin{figure}[ht!]
	\includegraphics[width=8cm]{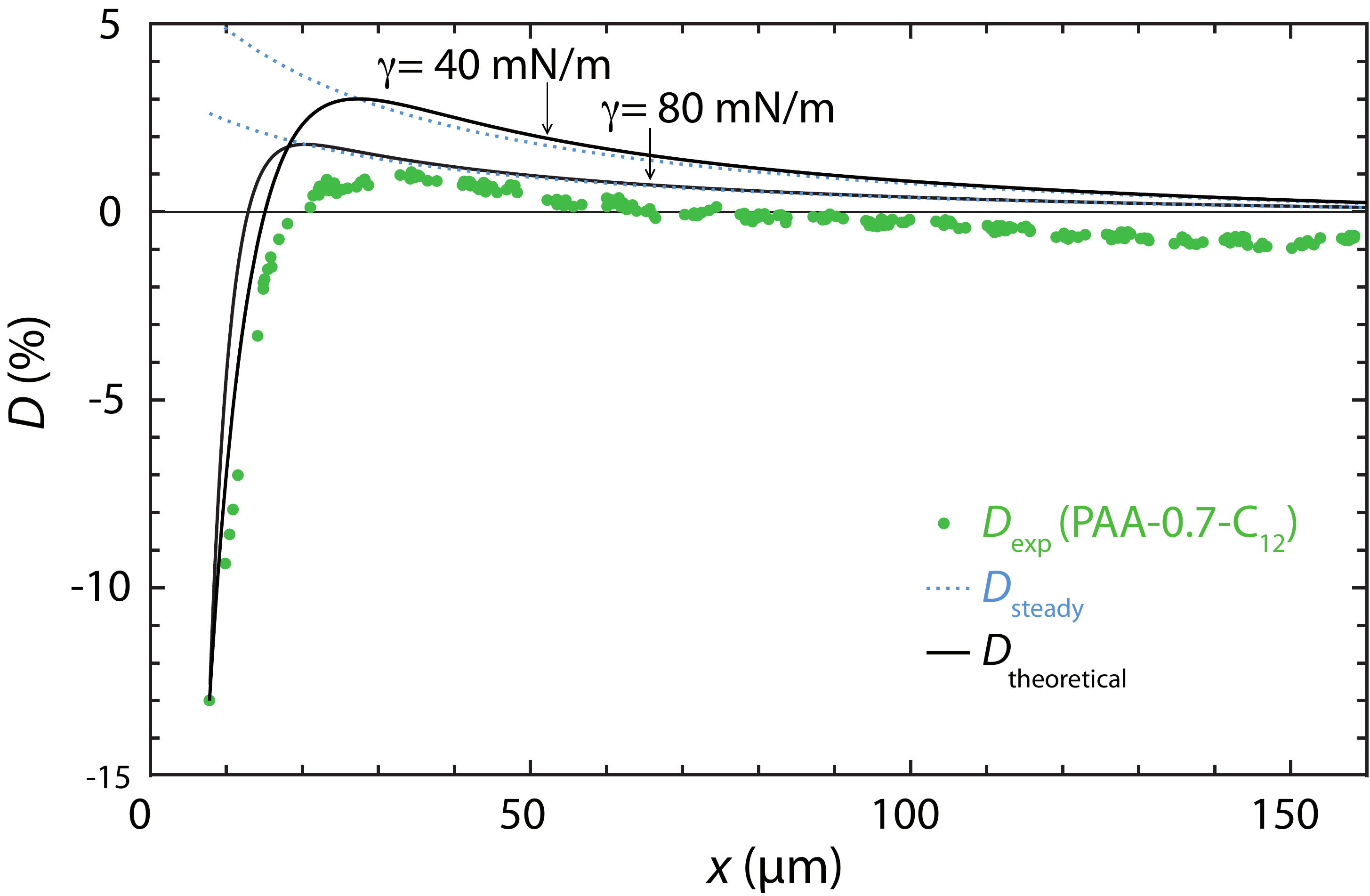}
	\caption{\label{1C12FitWrong}Deformation as a function of position for \paa{0.7}{12} monolayer capsules (see \fig{DeformationMonolayers}). The dashed and solid lines represent the predictions for the steady and theoretical deformations in the capillary model (Equations~\eqref{Dsteady} and~\eqref{TransientD}), with two test surface tensions as indicated.}
\end{figure}

We showed in our previous article that this model fits well the data obtained for the PMAA monolayer droplets\cite{Tregouet2018-PRF}. However, as shown in \fig{1C12FitWrong}, this simple capillary model can not fit the data obtained for the \paa{0.7}{12} capsules. More precisely, this model predicts that the deformation reaches positive values before returning to zero deformation. However the \paa{0.7}{12} returns to negative deformations at longer times. For the same reasons, the bilayer cases presented in \fig{DeformationPmaaLayers} can not be properly fitted (not shown) by the capillary model as they do not reach positive values of deformation. This means that, except for PMAA monolayers, our capsules cannot be considered as simple droplets with a homogeneous and constant interfacial tension.\\

However, we have several evidences that a capillary model is probably not sufficient to describe such membranes. Indeed, measuring the interfacial tension of \paa{0.7}{12} layers using the pendant-drop method and varying the area of the drops (by deflating them), we found \cite{Tregouet2016} that the surface tension decreases as the area decreases, similarly to the case of interfaces with insoluble surfactants \cite{Lucassen1972}. It is likely that the surface tension is not constant and homogeneous at the droplet interface as it is deformed. Such a membrane can be described by a surface dilational elastic modulus: $E_\gamma$ \cite{LUCASSEN2001}.\\
In the case of the PMAA/PVP, we found that these bilayers exhibit both shear and dilational moduli on the order of the surface tension value according to pendant-drop and shear-rheometer experiments\cite{LeTirilly2016a}. \\

In order to take into account these considerations, we alternatively consider these membranes as purely elastic ones.

\subsubsection{Elastic model}

In the case of purely elastic membranes, \textit{i.e.} in the absence of capillary effects and membrane viscosity, Barthes-Biesel and Rallison~\cite{BarthesBiesel1981} proposed a model to describe the relaxation of a slightly deformed  capsule toward its spherical shape. In such a case, the exponentional relaxation occurs in a steady liquid and is driven by the elasticity of the membrane, while the bulk viscosity of the droplet tends to slow down the dynamics. In a more general case, the deformation $D$ relaxes toward a reference deformation $D_{\textrm{ref}}$ following the equation:
\eqv{\dot{D}= \dfrac{1}{\tau\ind{el}} (D_{\textrm{ref}}-D)\label{ElasticRestoring}}
where the introduced time scale can take the two (note the possible signs) following values:
\eqv{\tau\ind{el\pm}= \dfrac{3(19 \lambda + 16)(2 \lambda +3)}{5(19 \lambda + 24 ) \pm \sqrt{5377 \lambda ^2+14256 \lambda + 9792}}\cdot \dfrac{\eta\ind{bulk}r}{3 G'}\label{ElasticTime0}}
with $G'$ the 2D shear elastic modulus (homogeneous to an energy per unit surface) of the membrane. In the simplest case, $D_{\textrm{ref}}$=0 and the capsules relax toward a spherical shape. As pointed out by Leclerc \etal~\cite{Leclerc2011}, the largest relaxation time rules the relaxation of the membrane which is then given by: $\tau\ind{el -}$.
Note that, using the same characteristic time, a variant of Equation~\eqref{ElasticRestoring} has been recently proposed by Gires \etal~\cite{Gires2016} for viscoelastic membranes, which also does not take into account the extensional flow (but only the shear flow).\\

Such models, by neglecting the extensional flow, can only predict a relaxation of the capsules from a negative deformation to a spherical shape and cannot account for a positive deformation as the one observed for \paa{0.7}{12}.

\subsubsection{Suggested model}

To account for all our experimental results, we suggest that both the interfacial tension and the elasticity of the membrane should be taken into account. The deformation should result from two contributions. 
The first one comes from the competition between the shear rate in the continuous phase which deforms the capsule and the surface tension which resists the deformation. This competition can be described dynamically by Equation~\eqref{TransientD}.
The second contribution corresponds to an elastic relaxation toward a reference shape $D_{\textrm{ref}}$, which is not necessarily zero. Such a dynamic is described by Equation~\eqref{ElasticRestoring}.\\

As previously mentioned, we point out that our membranes can be considered as elastic. Moreover, variations of the surface tension due to surface-excess variations are usually taken into the elastic effects together with the actual elasticity of the membrane. Both components contribute to the restoring forces and are defined by a global 2D effective modulus $E$ (homogeneous to an energy per unit surface). 

\begin{figure}[ht!]
\centering
\includegraphics[width=8cm]{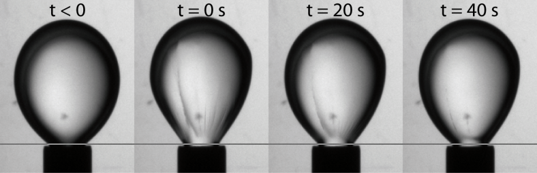}
\caption{\label{WrinklesPD}Wrinkling and relaxation of a PMAA/PVP membrane at the dodecane-water interface, in a pendant-drop apparatus: the membrane is compressed at time $t=0$, and then the inner volume is kept constant, while we observe the membrane relaxing over time. Scale: the outer diameter of the nozzle at the bottom is $1.2\, \un{mm}$.}
\end{figure}

Furthermore, we know thanks to interfacial rheology experiments that the polymer membranes do relax: the surface tension rises after compression and decreases after dilatation (not shown), and the shape of droplets covered by such membranes evolves over time after compression of the interface, as shown in Figure \ref{WrinklesPD}. This means that the polymer chains in the membrane relax.  At short time scales, the shape of the capsule should thus be a compromise between the initial shape (minimizing the energy for the elastic membrane that matured in the narrow channel) and the spherical shape (minimizing the area and hence the energy due to the homogeneous and isotropic surface tension). At long time scales, the membrane being viscoelastic, the polymer chains and thus the elastic stresses should relax and the droplet should retrieve its spherical shape. To include this potential membrane relaxation in our model, we allow for $D\ind{\textrm{ref}}(t)$ to be a function of time $t$ that relaxes towards zero over a viscoelastic polymeric time $\tau\ind{pol}$.

\begin{figure}[ht!]
	\includegraphics[width=8cm]{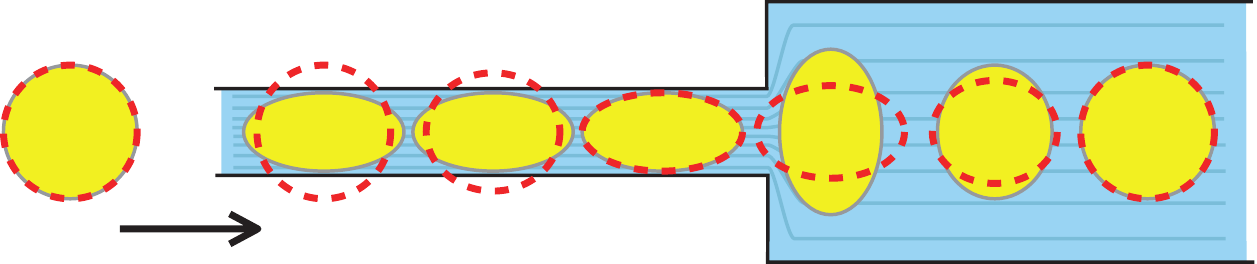}
	\caption{\label{DrefRelax} Schematic deformation of the capsule along its trajectory before the deformation chamber, with the corresponding reference deformation $D\ind{ref}$ (red dashed line). In the confining channel, when the elastic stresses relax, the reference deformation becomes the present deformation of the capsule. In the chamber, the reference deformation finally relaxes.}
\end{figure}

The initial reference deformation $D\ind{\textrm{ref}}(0)$ is thus comprised between $0$ (\textit{i.e.} the spherical shape of the droplet before the narrow channel) and the deformation of the capsule in the narrow channel, depending on the length of the latter and thus the maturation of the confined membrane. \fig{DrefRelax} illustrates the case where the narrow channel is long enough to have the restoring stresses totally relaxed before the capsule enters the wide chamber. In such a case, the reference deformation at the entrance of the wide chamber matches the actual deformation at that point. 

In our experiments, the capsules are confined in the narrow channel during a time $\tau\ind{conf}\simeq 10^{-2} \un{s}$. The long-channel approximation is thus valid if $\tau\ind{pol} \ll \tau\ind{conf}$, which has to be checked \textit{a posteriori}.

We would like here to insist on the distinction between the time $\tau\ind{pol}$ and the two other times $\tau\ind{ca}$ and $\tau\ind{el}$. The two latter times describe the relaxation of the capsule's shape according to a competition between a driving force (interfacial tension or elasticity) and the bulk viscosity. On the contrary, $\tau\ind{pol}$ describes the relaxation of the stresses within the membrane, and is thus related to the relaxation of the constitutive polymer chains which is the core of the membrane's viscoelasticity. The typical length scales associated to these two types of processes are thus totally different: $\tau\ind{ca}$ and $\tau\ind{el}$ describe the capsule as an object of a few tens of micrometers, while $\tau\ind{pol}$ describe the dynamics of the polymer chains, whose size is on the order of a few tens of nanometers. Accordingly, we expect $\tau\ind{pol}$ to be independent of the capsule's radius $r$, while we know from Equations \eqref{LiquidRelaxationTimeCapsule} and \eqref{ElasticTime0} that $\tau\ind{ca}$ and $\tau\ind{el}$ scale linearly with $r$.

To summarize, inspired by the pure cases above, we propose a more general model that simply assumes a superimposition of all the different contributions, and that couples them through the following equations (valid in the large chamber only):
\begingroup\makeatletter\def\f@size{8}\check@mathfonts
\eqcp{\label{MicrofluModel}}{\dot{D}(t)=\dfrac{1}{\tau\ind{ca}} (D\ind{steady}(t)-D(t)) + \dfrac{1}{\tau\ind{el}} (D\ind{ref}(t)-D(t))\label{MicrofluModela}}{\dot{D}\ind{ref}(t) = \dfrac{1}{\tau\ind{pol}} (0-D\ind{ref}(t)) \label{MicrofluModelb}}
\endgroup

The first term of the right-hand side of Equation \eqref{MicrofluModela} comes from Equation \eqref{TransientD}. It corresponds to a capillary relaxation towards an instantaneous steady state, and it is characterized by the time scale $\tau\ind{ca}$. The second term corresponds to an elastic relaxation towards an instantaneous reference state, and it is characterized by the time scale $\tau\ind{el}$ which includes the effects of different moduli: the ``liquid modulus" $E\ind{\gamma}$ describing the variations of the surface tension due to surface-excess variations, the 2D elastic compression modulus $K'$, and the 2D elastic shear modulus $G'$. We note $E$ the 2D effective modulus which is a combination of these three moduli, and that is expected to be close to the highest of these three moduli. We can thus write that:
\eqp{\tau\ind{el}\propto \dfrac{\eta\ind{bulk} r}{E} \label{tauRestor}}

Equation \eqref{MicrofluModelb} accounts for the relaxation of the stresses in the membrane, and the viscoleastic return to a spherical shape.
We would like to point out that a long relaxation time $\tau\ind{pol}$ of the polymers usually implies a high modulus in simple systems: for instance in the Lucassen model \cite{Lucassen1972}, a long relaxation time $\tau\ind{pol}$ is due to a low diffusivity. This implies a high value of the liquid modulus $E\ind{\gamma}$. Such systems would elastically relax over a short time $\tau\ind{el}$. In more complex systems, the link between $\tau\ind{pol}$ and $\tau\ind{el}$ can be less straightforward. This is in particular what we will see in the following. 

Solving the above coupled equations allows to predict a theoretical deformation of the capsules, with an interfacial tension and viscoelastic effects, in a viscous extensional flow. They will be confronted to the experimental data presented above, which will provide three fitting parameters, $\tau\ind{ca}$, $\tau\ind{el}$ and $\tau\ind{pol}$, and thus insights on the various mechanisms at play.

Finally, as a remark, we can recast Equation~\eqref{MicrofluModela} into an effective single-forcing process:
\eqv{\dot{D}(t)=\dfrac{1}{\tau\ind{forcing}} (D\ind{forcing}(t)-D(t))\label{MicrofluModelForcing}}
where:
\eqv{\tau\ind{forcing} = \Big({\tau\ind{ca}}^{-1}+{\tau\ind{el}}^{-1} \Big)^{-1}\label{ForcingTau}}
and:
\begingroup\makeatletter\def\f@size{8}\check@mathfonts
\eqp{D\ind{forcing}(t)= \tau\ind{forcing} \cdot \Big({\tau\ind{ca}}^{-1} \cdot D\ind{steady}(t)+{\tau\ind{el}}^{-1} \cdot D\ind{ref}(t)\Big) \label{ForcingD}}
\endgroup
This compact form highlights the universality of the model, that could be extended to other types of responses: more complex viscoelasticity, charge effects, poroelasticity...

\subsubsection{\paa{0.7}{12} monolayer: quickly-relaxing low modulus}

\begin{figure}[htp!]
	\includegraphics[width=8cm]{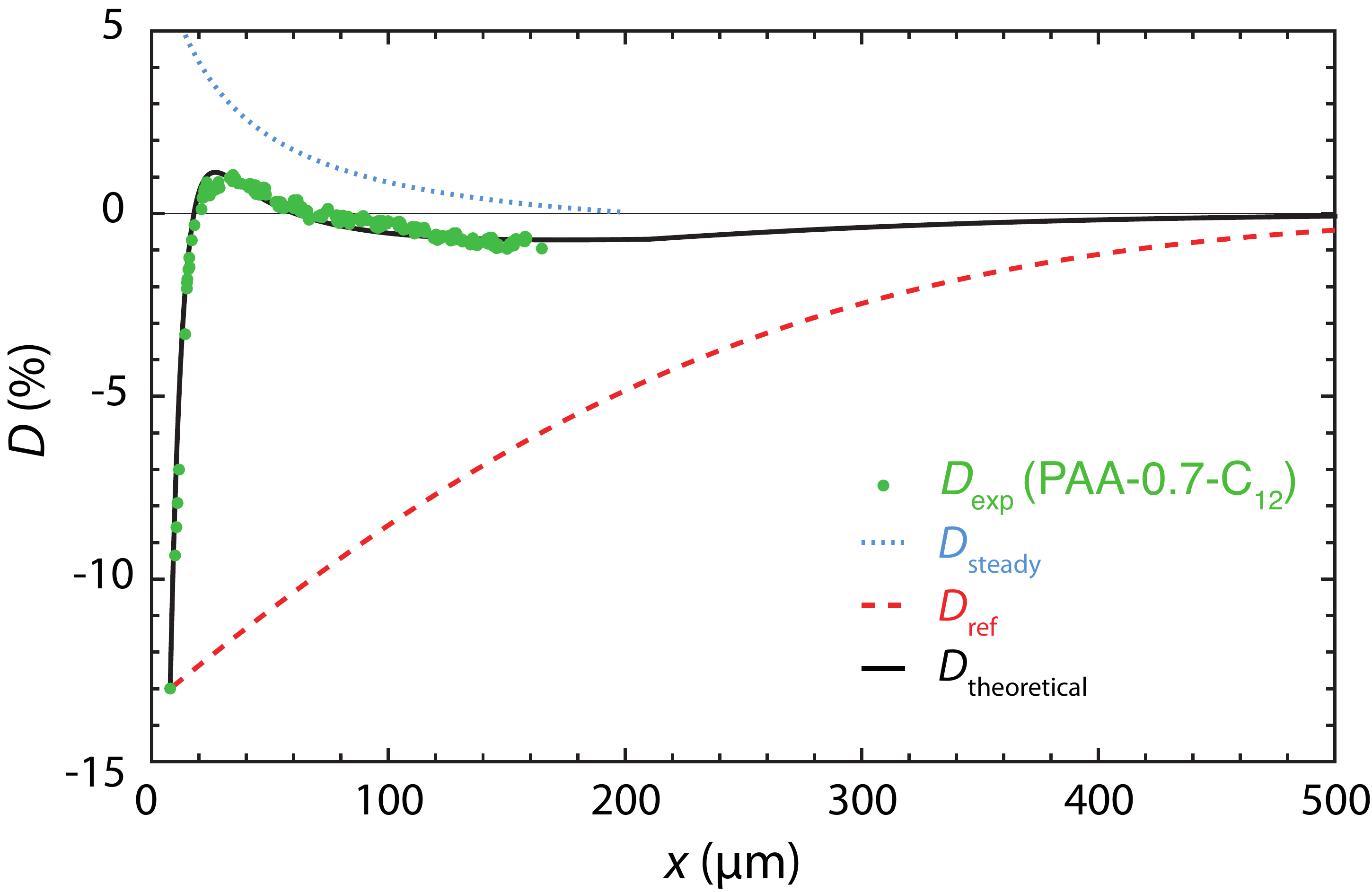}
	\caption{\label{Fit1C12}Deformation as a function of position for \paa{0.7}{12} monolayer capsules. The dashed lines represent the theoretical evolution of the two components : the steady deformation of the capillary model and the reference deformation of the elastic one. The solid line corresponds to the theoretical prediction of the global deformation given by Equation~\ref{MicrofluModel}. The fit of the data provides: $\tau\ind{ca}=3.03\pd{-5} \un{s}$ ($\gamma=35 \mnm$), $\tau\ind{el}=1.7\pd{-4} \un{s}$ and $\tau\ind{pol}=1.2\pd{-3} \un{s}$.}
\end{figure}

We use the general model presented in Equation~\eqref{MicrofluModel} to fit the experimental data for the \paa{0.7}{12} capsules. The fit shown in \fig{Fit1C12} provides: $\tau\ind{ca}=3.03 \pd{-5} \un{s}$ (\textit{i.e.} an interfacial tension of $35 \mnm$), $\tau\ind{el}=1.7\pd{-4} \un{s}$, and $\tau\ind{pol}=1.2\pd{-3} \un{s}$.
Pendant-drop experiments for an oil-water interface (not shown) indicated that the interfacial tension between mineral oil and  \linebreak \paa{0.7}{12} is about $25 \mnm$ and decreases under compression. Here, we find an interfacial tension that is higher than in pendant-drop experiments, but we have a fair order-of-magnitude agreement. Moreover, we observe that $\tau\ind{el}\gtrsim \tau\ind{ca}$, which we interpret as follows: the interfacial-tension variations (\textit{i.e} the liquid modulus) are low compared to the absolute value of the interfacial tension but not completely negligible. Finally, we find that $\tau\ind{pol} \ll \tau\ind{conf}$, which validates the long-channel approximation discussed above: at the end of the narrow channel, just before the capsules enter the wide chamber, all the elastic stresses have relaxed. Thus, the inital reference deformation $D\ind{ref}(0)$ corresponds to the actual shape of the capsules as they enter the chamber.

\subsubsection{\paa{0.7}{12}/PVP bilayer: slowly-relaxing high modulus}

\begin{figure}[ht!]
	\includegraphics[width=8cm]{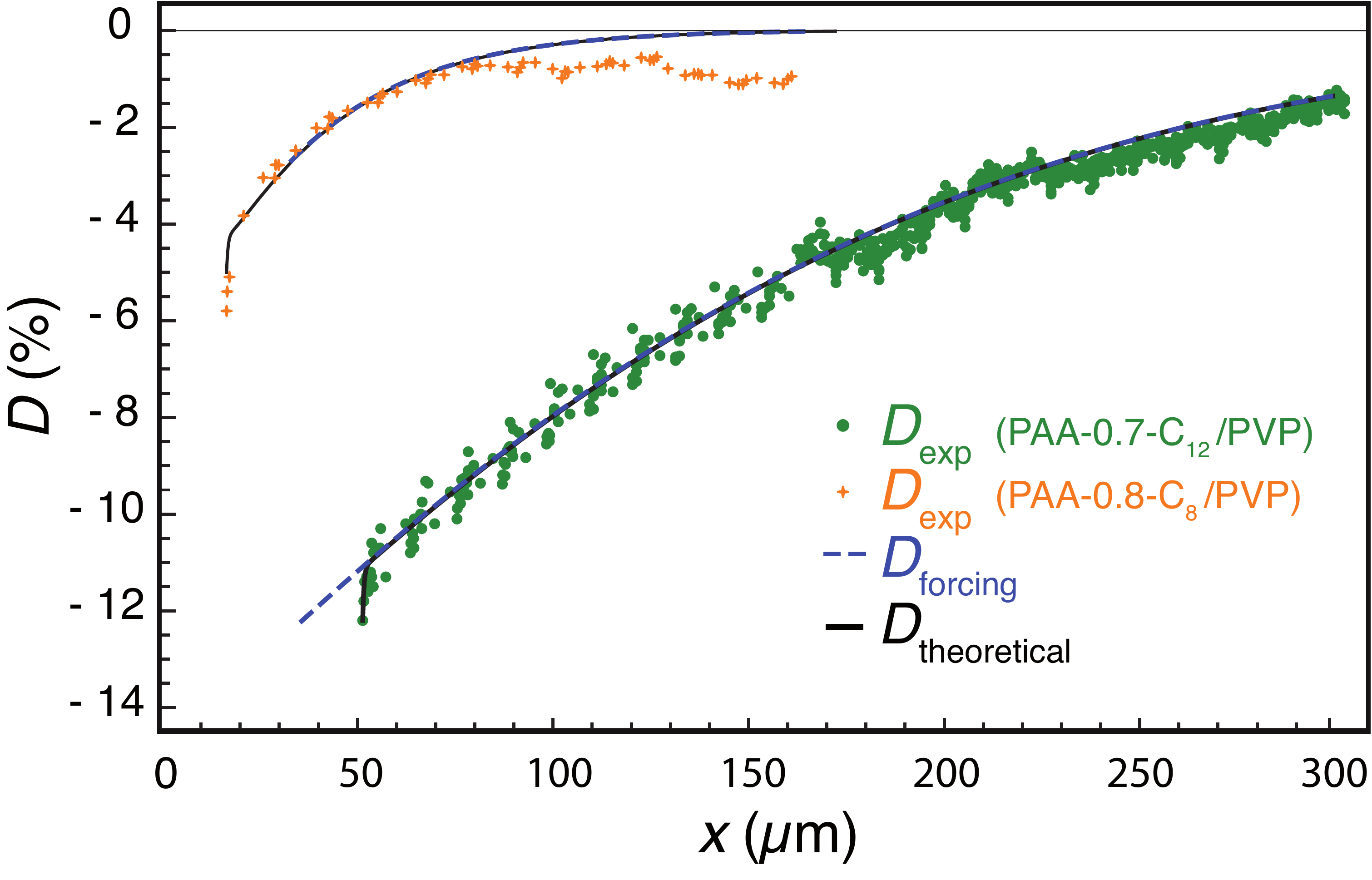}
	\caption{\label{1C12-1C8}Deformation as a function of position for \paa{0.7}{12}/PVP and \paa{0.8}{8}/PVP bilayer capsules. The dashed and solid lines represent the predictions for the forcing and theoretical deformations in the general model. For the \paa{0.7}{12}/PVP bilayer, the best fit provides:  $\tau\ind{ca}=3.25\pd{-5} \un{s}$ (\mbox{$\gamma=35 \mnm$}), $\tau\ind{el}=3\pd{-6} \un{s}$, and $\tau\ind{pol}=1.2\pd{-3} \un{s}$. For the \paa{0.8}{8}/PVP bilayer, the best fit provides: $\tau\ind{ca}=3.67\pd{-5} \un{s}$ (\mbox{$\gamma=35 \mnm$}), $\tau\ind{el}=3\pd{-6} \un{s}$, and $\tau\ind{pol}=2.5\pd{-4} \un{s}$.}
\end{figure}

We apply the same analysis to the capsules covered by a \paa{0.7}{12}/PVP bilayer, as presented in \fig{1C12-1C8}. The best fit gives $\tau\ind{ca}=3.25 \pd{-5} \un{s}$, and thus (we recall here that $\tau\ind{ca}$ also depends on the capsule's radius, which slightly varies between the different experiments) the same interfacial tension ($\gamma=35 \mnm$) as for the \paa{0.7}{12} monolayer. Moreover, we find the same polymeric relaxation time ($\tau\ind{pol}= 1.2 \pd{-3} \un{s}$) as in the case of the \paa{0.7}{12} monolayer, but a significantly lower elastic time $\tau\ind{el}=3.0 \pd{-6} \un{s}$. According to Equation~\eqref{tauRestor}, this implies that the effective elastic modulus $E$ is significantly higher for the bilayer than for the monolayer, which is consistent with the added interactions between \paa{0.7}{12} and PVP chains.

The value obtained for the polymeric relaxation time ($\tau\ind{pol}\simeq 1 \un{ms}$) is also coherent with the pendant-drop experiments reported previously \cite{LeTirilly2016a}. Therein, there was no evidence of any surface-tension inhomogeneity, or anisotropy, during compression, neither for the \paa{0.7}{12}/PVP bilayers nor for the \paa{0.7}{12} monolayers, which indicates that the inner relaxation times are shorter than the typical time scale of such experiments (on the order of $\sim 1 \un{s}$). 

Furthermore, we interpret the concordance of the polymer relaxation time $\tau\ind{pol}$ between the monolayer and the bilayer as a consequence of the predominance of the alkyle graft anchoring on the dynamics of the interface, as shown in a previous work \cite{LeTirilly2016a}. A second evidence for the importance of graft anchoring is the comparison with an analogous experiment performed on \paa{0.8}{8}/PVP bilayers, also presented in \fig{1C12-1C8}. We observe that the relaxation of such bilayers is much faster than the relaxation of \linebreak \paa{0.7}{12}/PVP bilayers. In this simple picture, shorter alkyle grafts, and thus lower anchoring energy, leads to faster relaxation. 
In such a case, if the relaxation of the bilayers was instead controlled by the moduli only (and thus $\tau\ind{el}$), we would expect a lower elastic modulus for the \paa{0.8}{8} compared to the \paa{0.7}{12} as these chains are less strongly anchored to the interface\cite{Barentin1998}. In such case, according to Equation~\eqref{tauRestor}, we would observe a faster relaxation for \paa{0.7}{12}/PVP than for \paa{0.8}{8}/PVP -- which is opposite to what we measured. This comparison between \paa{0.8}{8}/PVP and \paa{0.7}{12}/PVP thus indicates that the relaxation time which rules the overall relaxation of those capsules is the one for polymer relaxation $\tau\ind{pol}$, and that graft anchoring is a key factor of the underlying polymer dynamics.

\subsubsection{PMAA/PVP bilayer: barely-relaxing high modulus}

Unlike the \paa{0.7}{12}/PVP bilayers, the PMAA/PVP bilayers show wrinkles during several seconds after compression in pendant-drop experiments, as presented in \fig{WrinklesPD}. Accordingly, we expect to have in this case $\tau\ind{pol} > 1 \un{s}$, which has two implications. First, this time scale is much longer than the time during which the capsules are confined in the narrow channel $\tau\ind{conf}$. As a consequence, we assume that such capsules do not have enough time to undergo any viscoelastic maturation process within the narrow channel, and that the reference deformation $D\ind{ref}(0)$ at the entrance of the wide chamber is close to zero (because of the memory of the non-deformed state before confinement). Secondly, the expected value of $\tau\ind{pol}$ is so high that it can be considered as infinite when compared to the typical time scales of our microfluidic measurements, which is about $1 \un{ms}$. Therefore, we fix the value of the reference deformation $D\ind{ref}$ to a (almost zero) constant at all times.

\begin{figure}[ht!]
	\includegraphics[width=8cm]{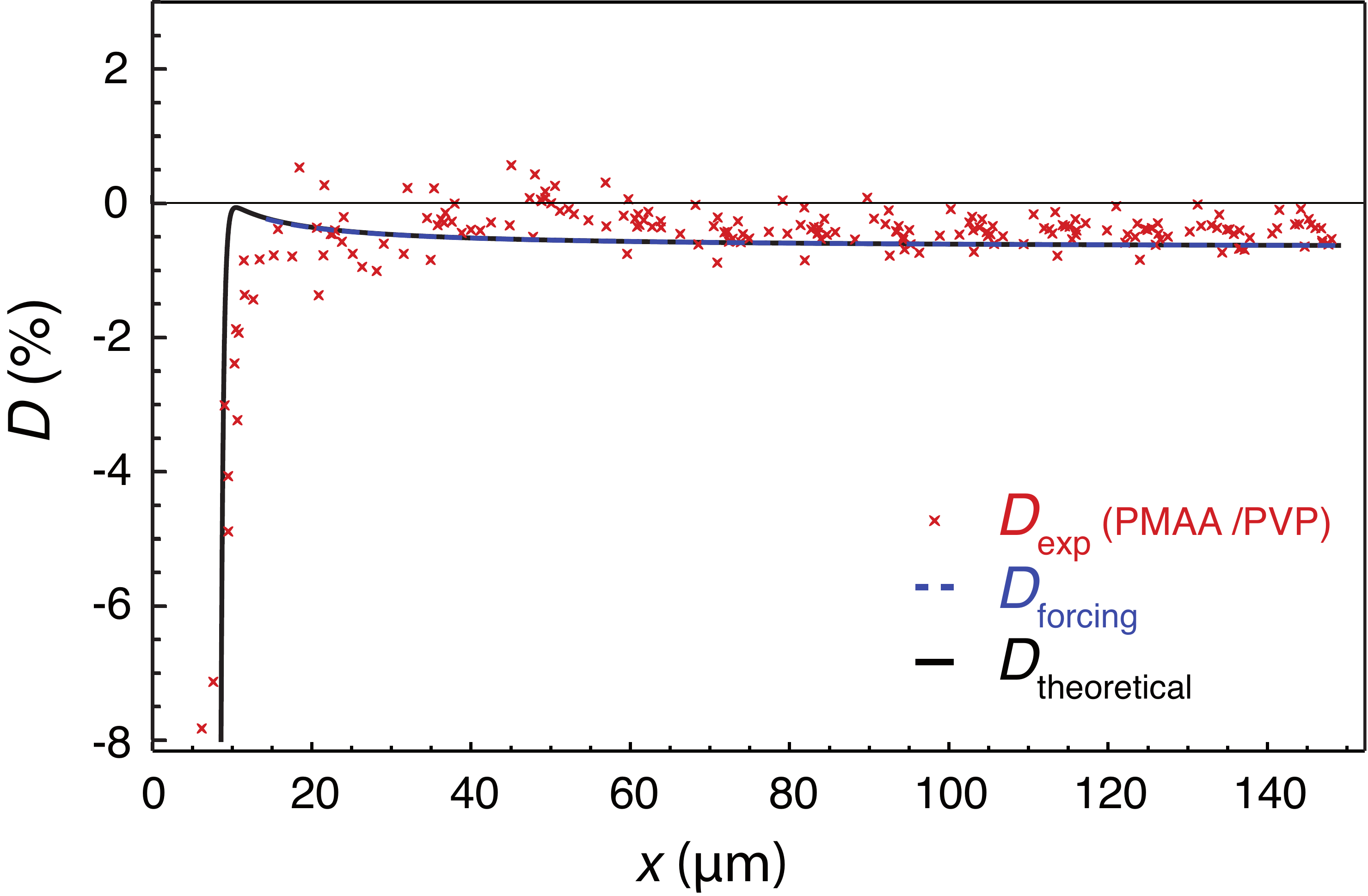}
	\caption{\label{FitPmaaPvp}Deformation as a function of position for PMAA/PVP bilayer capsules. The dashed and solid lines represent the predictions for the forcing and theoretical deformations in the general model. The best fit provides: $\tau\ind{ca}=2.8\pd{-5} \un{s}$ ($\gamma=40 \mnm$)  and $\tau\ind{el}=3\pd{-6} \un{s}$. We assumed $\tau\ind{pol}\gg 10^{-3} \un{s}$.}
\end{figure}

The result of the fit is presented in \fig{FitPmaaPvp}. We find the same interfacial tension $\gamma= 40 \mnm$ as for the PMAA monolayer~\cite{Tregouet2018-PRF}, and an elastic time $\tau\ind{el}=3\pd {-6} \un{s}$ similar to the elastic time of \paa{0.7}{12}/PVP. 

\subsection{Discussion}

We summarize in Table~\ref{TableTimes} all the results obtained by fitting the experiments to the general model. The effective 2D moduli $E$ are calculated from $\tau\ind{el}$ according to Equation~\eqref{ElasticTime0}. 
\begin{table*}[ht!]
	\small
	\begin{center}
    \caption{\label{TableTimes}Best-fit parameters extracted from the comparison between all the experimental data and the general model.}
\begin{tabular}{|l|c|c|c|c|c|c|}
			\hline
			$\;$ & $\;$ & $\;$ & $\;$ & $\;$ & $\;$  \\ [-2 mm]
			$\;$ &  $\gamma \;(\mathrm{mN/m})$ & $\tau\ind{ca}\;(\mathrm{s})$ & $\tau\ind{el} \;(\mathrm{s})$ & $\tau\ind{pol} \;(\mathrm{s})$ & $E \;(\mathrm{mN/m})$  \\ [2 mm] \hline
			$\;$ & $\;$ & $\;$ & $\;$ & $\;$ & $\;$  \\ [-2 mm]

PMAA \cite{Tregouet2018-PRF}& $40$ & $2.33\pd{-5}$ & $-$ & $-$ & $-$  \\ [3 mm]
			\paa{0.7}{12}  & $35$ & $3.03\pd{-5}$ & $1.7\pd{-4}$ & $1.2\pd{-3}$ & $ 1.3\pd{1}$ \\ [3 mm]
			\paa{0.7}{12}/PVP  & $35$ & $3.25 \pd {-5}$ & $3\pd{-6}$ & $1.2\pd{-3}$ & $ 6.9\pd{2}$ \\ [3 mm] 
			\paa{0.8}{8}/PVP  & $35$ & $3.67\pd{-5}$ & $3\pd{-6}$ & $2.5\pd{-4}$ & $ 7.7\pd{2}$ \\ [3 mm]
			PMAA/PVP  & $40$ & $2.8 \pd {-5}$ & $3\pd{-6}$ & $\gg 10^{-3}$ & $ 6.9\pd{2}$ \\ [2 mm] \hline
		\end{tabular}
	\end{center}
	\normalsize
\end{table*}
The orders of magnitude of the obtained moduli are in fair agreement with what we expect from macroscopic measurements, when available. In the case of \paa{0.7}{12}, where the modulus is a liquid dilational modulus $E\ind{\gamma}$, if there is no desorption, we can show that $E\ind{\gamma}=\Pi$, where $\Pi$ is the surface pressure appearing in the state equation of Leclerc \textit{et al.}~\cite{Daoud1999} (cf Supplementary Information). The surface pressure of \paa{0.7}{12} at the mineral oil-water interface is $\Pi \simeq 10 \mnm$, which leads to $E\ind{\gamma}\simeq 10 \mnm$, in good agreement with what we obtained in our experiments. For the \paa{0.7}{12}/PVP bilayer, no measurement of the elastic modulus has been performed at time scales smaller than $\tau\ind{pol}\simeq 1 \un{ms}$, and consequently we have no reference for comparison. In contrast, the value extracted from the experiment with PMAA/PVP is close to the value measured in the interfacial rheometer \cite{LeTirilly2015,LeTirilly2016a}. 

In summary, the general model described in Equation~\eqref{MicrofluModel} offers both a good understanding of the relaxation dynamics of various capsules in a microfluidic device, and a satisfactory agreement with calibrated experiments concerning the extracted physical parameters. The strong influence of the ratio $\tau\ind{pol}/\tau\ind{conf}$ between the polymer and confinement time scales (which \textit{e.g.} controls the differences observed between the \paa{0.7}{12}/PVP and PMAA/PVP systems) further suggests that it might be interesting to perform new sets of experiments by varying the length of the narrow channel. This would be a way to extract more precisely $\tau\ind{pol}$, and to characterize further the underlying molecular mechanisms for relaxation -- a key ingredient for understanding and tuning finely the membrane properties. 

\section{Conclusion}

Through the use of a sharp microfluidic-channel extension we measured the shape relaxation of capsules with different kinds of polymer membranes. We observed various behaviors depending of the viscoelastic parameters of these membranes. We proposed a general model to understand this rich set of behaviors, arising from the relative importances of the outer fluid viscosity, and the interfacial tension, elastic modulus, and/or viscous modulus of the membranes. The comparison of the experiments with the model predictions allows us to rationalize all the experimental results in a unified framework. Based on this work, by screening different confinement times, we envision the possibility of probing finely, and \textit{in situ}, the rheological and molecular behaviours of capsules and cells within microfluidic plateforms.

\subsubsection*{Acknowledgements}

This work was financially supported by ANR JCJC INTERPOL. The authors thank Mamisoa Nomena and Samuel Poincloux for their precious help and advice in this work. They also thank Oliver B\"aumchen, Ingmar Polenz, Jean-Christophe Baret, Julien Dupr\'e de Baubigny, Nad\`ege Pantoustier and Patrick Perrin for fruitful discussions.

\balance

\bibliography{MicrofluComplex}
\bibliographystyle{unsrt}

\end{document}